%% file: ms.tex
\documentclass[apj]{emulateapj}
\usepackage[usenames]{color}
\usepackage{soul}

\shorttitle{MASS TRANSFER IN BINARIES. II.}
\shortauthors{LAJOIE \& SILLS}

\begin{document}

\title{Mass Transfer in Binary Stars using SPH. II. Eccentric Binaries}  

\author{Charles-Philippe Lajoie \& Alison Sills}
\affil{Department of Physics and Astronomy, McMaster University,
  \\Hamilton, ON L8S 4M1, Canada}
\email{lajoiec@physics.mcmaster.ca,\\asills@mcmaster.ca}

\begin{abstract}
  Despite numerous efforts to better understand binary star evolution,
  some aspects of it remain poorly constrained.  In particular, the
  evolution of eccentric binaries has remained elusive mainly because
  the Roche lobe formalism derived for circular binaries does not
  apply.  Here, we report the results of our Smoothed Particle
  Hydrodynamics simulations of mass transfer in eccentric binaries
  using an alternate method in which we model only the outermost
  layers of the stars with appropriate boundary conditions.  Using
  this technique, along with properly relaxed model stars, we
  characterize the mass transfer episodes of binaries with various
  orbital parameters. In particular, we show that these episodes can
  be described by Gaussians with a FWMH of $\sim0.12$
  P$_{\textrm{orb}}$ and that the peak rates occur after periastron,
  at an orbital phase of $\sim0.58$, independently of the eccentricity
  and mass of the stars.  The accreted material is observed to form a
  rather sparse envelope around either or both stars. Although the
  fate of this envelope is not modeled in our simulations, we show
  that a constant fraction ($\sim5\%$) of the material transferred is
  ejected from the systems.  We discuss this result in terms of the
  non-conservative mass transfer scenario. We suggest our results
  could be incorporated in analytical and binary population synthesis
  studies to help better understand the evolution of eccentric
  binaries and the formation of exotic stellar populations.
\end{abstract}

\section{INTRODUCTION}
\label{sect:intro}
The study of exotic stellar populations (e.g.\ blue stragglers,
low-mass X-ray binaries, helium white dwarfs) requires the
understanding of close binary evolution. In turn, the study of binary
evolution involves the use of many physical mechanisms occurring over
dynamical, thermal, and nuclear timescales, which quickly render the
problem at hand complicated.  The analytical tools generally used to
study binary stars have difficulty resolving all of these timescales;
instead, they usually incorporate only some of the mechanisms or rely
on analytical approximations. On the one hand, stellar evolution codes
can evolve single stars over many billion years while taking into
account convective mixing and different nuclear reactions
networks. Only recently have they been used to evolve binary stars,
although most of them are still in one dimension and only a handful
use two dimensions \citep{han2000,deupree2005}.  These binary
evolution codes, along with population synthesis codes, which evolve
many millions of stars at once over nuclear timescales
\citep{pzv1996,hurley2002,ivanova2005}, rely on analytical
prescriptions for the mass transfer and accretion rates.  None of
these techniques actually models the mass transfer itself as it often
occurs on timescales that are too short. On the other hand,
hydrodynamics is well suited for purposes such as mass
transfer. However, it can be difficult to incorporate physical
ingredients such as convective mixing, magnetic fields, radiative
transfer or nuclear reactions. Hydrodynamics is also usually not
designed to evolve stars over long periods of time. These difficulties
therefore render the modeling of the long-term hydrodynamical
evolution of interacting binaries rather challenging.

The ideal case of circular orbits and conservative mass transfer has
been studied intensively over the years. Seminal work by
\citet{morton1960}, \citet{pacz1965,pacz1971} and \citet{pacz1972},
among others, on mass transfer and its consequences on the stars and
the orbital parameters have opened the way to a more quantitative
study of binary evolution.  \citet{iben1991} and \citet{iben1993} have
more recently laid out the overall evolutionary paths of many
different binary populations and explained the formation scenarios of
many exotic objects. But, from a theoretical point of view, a detailed
description of some aspects of close interactions are still
lacking. Of particular interest are the rate at which mass is
transferred from one star to the other, the amount of mass accreted by
the secondary stars, and the degree of mass loss from these systems,
if any. To date, these quantities have usually been either
approximated from theoretical estimates or arbitrarily fixed.  But
because these quantities are critical for understanding the long-term
evolution of binary stars, as they are some of the mechanisms that
drive the change of orbital separation and ultimately dictate the fate
of binaries, it is important to get better estimates for the more
realistic, non-idealized cases.  In particular, recent studies
(\citealt{sepinsky2007a}, 2007b, 2009) have suggested that eccentric
binaries may evolve differently when compared to circular
binaries. Given that a non-negligible fraction ($\sim 20\%$) of
interacting binaries have eccentric orbits
\citep{petrova1999,raguzova2005}, this could arguably modify the
formation rates and total numbers of binary and exotic populations in
synthesis models.  Therefore, a better understanding and calibration
of the mass transfer rates, the degree of mass loss from binary
systems, and the accretion process are needed in order to get a
better, more realistic picture of the exotic stellar populations.

In this paper, we present the results of hydrodynamical simulations of
mass transfer in eccentric binaries using a specifically designed SPH
code (\citealt{lajoie2010I}; hereafter Paper I).  We model two
different binary systems with various eccentricities and semi-major
axes and concentrate on how and when mass transfer is initiated, as
well as how much mass is accreted by the companion star and/or lost
from the system.

\section{Brief theory of eccentric binary systems}
\label{sect:theory}
Although most short-period binaries are on circular orbits, many
relatively close binaries also have eccentric orbits in which mass
transfer can occurs only close to periastron. Such episodes of mass
transfer are usually not taken into account in binary population
synthesis studies since rapid circularization at the onset of mass
transfer is often assumed.  To some extent, such episodes of mass
transfer could modify the general picture of exotic star populations,
especially in dense clusters where the formation of eccentric binaries
through captures is more likely.  However, most of the theoretical
background generally used applies only to circular orbits, and one
must rely on other approximations to estimate the rate of mass
transfer in eccentric binaries, as we now discuss.

\subsection{Roche lobe and equipotentials}
\label{sect:roche}
Using analytics to investigate the equations of motion in binary
systems, \citet{sepinsky2007a} showed that eccentric binaries
undergoing mass transfer can behave quite differently when compared
with circular binaries.  Indeed, the authors found that the Roche lobe
radius (denoted $R_L$) can be smaller than the circular case by more
than $20 \%$ for binaries with mass ratios close to unity and rotating
faster than the orbital velocity at periastron.  The reverse is also
true, as binaries rotating slower than the orbital velocity at
periastron can have a Roche lobe radius $\sim 10\%$ \textit{larger}
than the circular case. Moreover, depending on the degree of
asynchronicity and eccentricity, the geometry of the equipotential
surfaces is found to change significantly and allow for some mass to
be ejected from the system through the $L_1$ point.
\citet{sepinsky2007a} found that the usual Roche lobe can sometimes
open up around the secondary star, allowing for some mass loss through
the $L_1$ point, and that the potential at the $L_3$ point can
sometimes be only slightly larger than that at the $L_1$ point, also
allowing for some more mass loss (see Figure 7 of
\citealt{sepinsky2007a}).  Although we expect some mass to be ejected
from the system, the total amount lost is hard to estimate and authors
have generally used some parameterizations to study the effect of
non-conservative mass transfer on the evolution of binaries
(\citealt{sepinsky2009}; hereafter, SWKR09).  These results suggest
that eccentric and asynchronous binaries are likely to undergo mass
transfer at earlier phases of their life (compared to circular
binaries) and that the latter is most likely non-conservative.  Note
that \citet{dermine2009} finds similar changes to the shape of
equipotentials when considering the effect of radiation pressure in
\textit{circular} and \textit{synchronized} binaries.

These recent works emphasize that the classical Roche model is not
adequate in most instances.  The addition of realistic physical
ingredients (e.g.\ asynchronism, eccentricity, radiation pressure) in
the models of binary stars modifies the structure of equipotentials.
The Roche lobe model derived for circular orbits therefore does not
apply.

\subsection{Secular evolution} 
\label{sect:secularevol} 
Based on their previous results, \citet{sepinsky2007b} (hereafter,
SWKR07) and SWKR09 went on to study the secular evolution of eccentric
binaries undergoing mass transfer, with the assumptions of
instantaneous mass transfer ($\dot{\textrm{M}}_0=10^{-9}$
M$_{\odot}$/yr) centered at periastron and both conservative and
non-conservative mass transfer.  The authors found that depending on
the mass ratio and eccentricity, the secular changes of orbital
separation and eccentricity can be positive or negative, and can occur
on timescales ranging from a few million years to a few billion years.
Moreover, these timescales can, in some cases, be comparable to the
orbital evolution timescales due to tidal dissipation, which can be
additive or competitive.  Based on these findings, the authors suggest
that the usual rapid circularization assumption is not always
applicable and, in some cases, very unlikely.  Finally, SWKR09
conclude that relaxing the assumption of conservative mass transfer
does not change the overall conclusions of their previous work
(SWKR07).  The rates of secular evolution for $a$ and $e$ found by
SWKR07 and SWKR09 are directly proportional to the assumed mass
transfer rate. However, this can be hard to constrain with analytical
prescriptions only. Indeed, when mass transfer occurs periodically,
binaries can remain on eccentric orbits for long periods of time,
making the Roche lobe radius and the mass transfer rate difficult to
determine, as the latter depends on the degree of overflow ($\Delta
R=R_*-R_L$, where $R_*$ is the radius of the star).

\subsection{Previous simulations of mass transfer}  
One way to better estimate the rates of mass transfer is by using
hydrodynamical simulations. Only a handful of such simulations have
been done to this day. Despite usually not being suited for long,
thermal- or nuclear-timescale studies, hydrodynamical simulations can
be useful for understanding transient phenomena and episodes of
dynamical mass transfer.

Only a few authors have investigated the hydrodynamics of eccentric
binaries. \citet{regos2005} (see also \citealt{layton1998}) studied
the shape of the equipotential surfaces in eccentric binaries using
both analytical and numerical (SPH) approaches. Their findings agree
with those of \citet{sepinsky2007a} in that mass transferred through
the $L_1$ point close to periastron passages may leave the system (as
well as through the $L_2$ point). However, their estimates for the
Roche radius are larger and similar to the Roche lobe radii for the
circular and synchronized case.  Interestingly, the authors also study
the onset of mass transfer along the orbit for one binary and
different eccentricities.  The low resolution of these simulations
(10,000 particles), however, does not allow for accurate mass transfer
rate determinations.

\citet{church2009} have partially circumvented this problem using an
innovative SPH technique for modeling mass transfer in cataclysmic
variables, where the least massive star is losing mass to a compact
white dwarf.  With the aim of getting better estimates of mass
transfer rates, their innovative approach allows for high
\textit{mass} resolution in the outer parts of the star and therefore
for the resolution of low mass transfer rates.  Despite using a
relatively low number of particles ($\sim40,000$), most of the stars'
mass is contained in a few very massive particles, allowing for the
outer particles to have relatively low masses.  By varying the
eccentricity and periastron distances for one particular mass ratio
($q_2=0.6$), the mass transfer rates they obtain show qualitative
behaviour in agreement with the photospheric mass transfer rate
predicted by \citet{ritter1988}.  \citet{edwards1987} performed
grid-based hydrodynamics calculations of polytropic semi-detached
systems and compared the mass transfer rates to analytical estimates,
with which they find good agreement. However, their simulations only
modeled a small rectangular box close to the $L_1$ point and did not
encompass either the donor or the accretor and did not assess whether
mass was lost from the system.

Finally, in some simulations to date, the accreting star is not
realistically modeled but rather often modeled as a point mass or with
surface boundary conditions. These simplifications prevent from
drawing any quantitative conclusions regarding the accretion process.
Moreover, as pointed out by \citet{sills1997}, the use of polytropes
instead of realistic models may lead to significantly different
internal structures for collision products, which may arguably be
applicable to interacting binaries.  Therefore, more work remains to
be done in order to better understand how mass transfer operates and
affects the evolution of eccentric binary systems.

\section{Computational Method}
\label{sect:method}
For a more realistic modeling of hydrodynamical mass transfer, it is
better to use hydrodynamics techniques since they can easily be
adapted to model binary systems in three dimensions and physically
follow the transfer of mass from one star to the other.  Here, we use
an SPH code based on the version of \citet{benz1990} and
\citet{bate1995} with a recent updated treatment of boundary
conditions specifically designed to model boundary stars presented in
Paper I. We model our stars from theoretical profiles obtained from
our stellar evolution code (YREC; \citealt{guenther1992}) and
distribute SPH particles on an hexagonal lattice while iteratively
assigning particle masses so that the density profile matches that
from our stellar evolution code.  Binaries are relaxed in their own
gravitational force (and centrifugal force) prior to the start of the
mass transfer simulations (see Paper I).  Using our treatment of
boundary conditions, we replace the inner particles with a central
point mass and model only the outermost layers of the stars.  The
location of the boundary is, at this point, arbitrary but should be
placed at least a few smoothing lengths from the surface.  Use of our
boundary conditions allows for better spatial and mass resolutions in
the mass transfer stream as well as the use of less CPU time.  Note
that each particle's smoothing length is also consistently evolved in
time, following the prescription of \citet{benz1990}, allowing for a
better spatial resolution in regions of high density. Finally, we use
Monaghan's viscosity \citep{monaghan1989} with $\alpha=1.0$ and
$\beta=2.0$ along with an adiabatic equation of state of the form
$P=(\gamma-1)\rho u$, where $\gamma=5/3$, $\rho$ is the density and
$u$ the internal energy (per unit mass).

\section{Mass Transfer in Eccentric Binaries} 
We now present the results of our simulations of mass transfer for two
different binary systems and discuss the overall behaviours observed
in our simulations.  In particular, we are interested in the mass
transfer rates and properties involved in such close interactions.  We
have modeled binary systems with stars of different masses, semi-major
axes, and eccentricity. The different orbital parameters modeled for
both system are summarized in Table \ref{tab:models} along with some
preliminary results.

\subsection{0.80 M$_{\odot}$ + 0.48 M$_{\odot}$}
\begin{figure}
  \begin{center}
    \includegraphics[angle=-90.,scale=0.7]{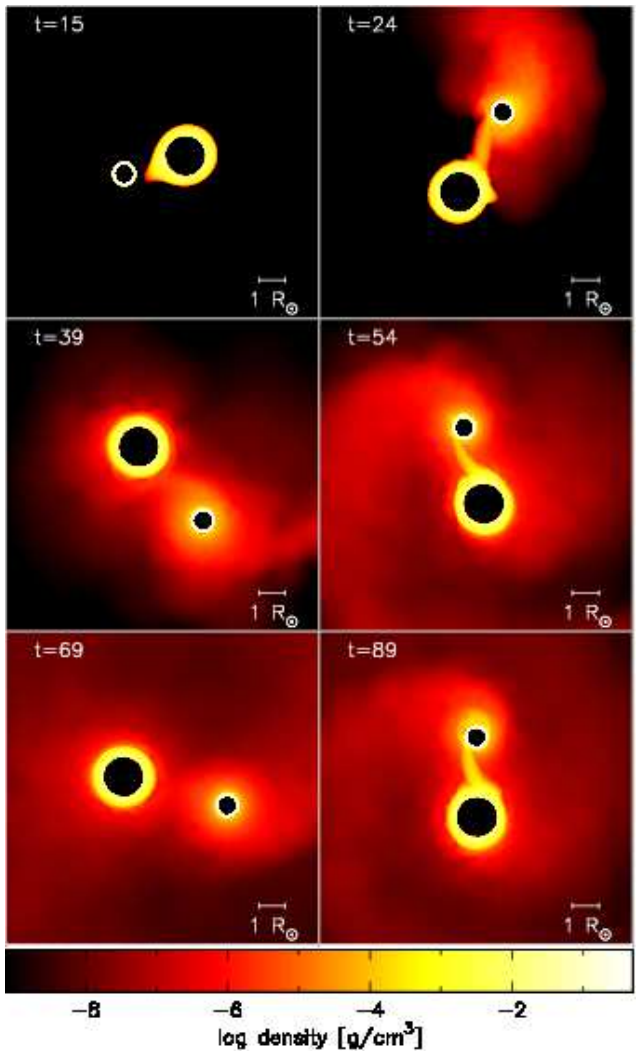}
    \caption {Logarithm of the density in the $XY$-plane for the
      $0.80+0.48$ M$_{\odot}$ binary and $e=0.25$. Each image is 12
      R$_{\odot}$ by 12 R$_{\odot}$ and the central point masses are
      not shown. The time is shown in units of the dynamical timescale
      ($\tau_{dyn}$) and the orbital period is $\sim 32$
      $\tau_{dyn}$.}
    \label{fig:XY_R006}
  \end{center}
\end{figure}

Our first model consists of a low-mass binary system representative of
the turn-off mass of globular clusters.  The initial separation of
this binary system, at apastron, is set to 4 R$_{\odot}$ such that the
stars do not initially overflow their Roche lobe.  The boundary is set
at 0.8 R$_{\odot}$ for the 0.8-M$_{\odot}$ star and 0.35 R$_{\odot}$
for the 0.48-M$_{\odot}$ star, both corresponding to $\sim$ 75$\%$ of
the stars' radii.  At this radius, most of the mass of the stars is
encompassed within the central point mass. The total number of
particles is $\sim 600,000$ and the total mass of SPH particles is
$\sim4\times 10^{-3}$ M$_{\odot}$ and $\sim3\times 10^{-2}$
M$_{\odot}$ for the 0.8- and 0.48-M$_{\odot}$ stars respectively.
Using \verb|SPLASH| \citep{splash2007}, a publicly available
visualization tool for SPH simulations, we show in Figure
\ref{fig:XY_R006} the logarithm of the density, in the $XY$ plane, for
the case with $e=0.25$.  The time is shown in units of the dynamical
time ($\tau_{dyn}= \sqrt{R^3_{\odot}/GM_{\odot}}\simeq 0.5$ hour) and
the orbital period corresponds to $\sim 32$ $\tau_{dyn}$.  Each image
is 12 R$_{\odot}$ by 12 R$_{\odot}$, and the density scale ranges from
10$^{-10}$ g cm$^{-1}$ (dark) to 1 g cm$^{-1}$ (white).
The 0.8-M$_{\odot}$ star is the larger of the two stars and the large
density contrast between the two stars is obvious from these
plots. Mass transfer occurs only periodically, close to periastron,
and shuts off when the stars are further apart. Moreover, the
secondary is retaining some of the transferred mass, forming an
envelope, whereas the primary does not seem to be affected strongly
from losing mass.  The density of the accreted material around the
secondary is much lower than that of the secondary's surface layer and
this may have some implications for the long-term accretion of this
material (see $\S$ \ref{sect:accretion}).  Some mass is also lost from
the secondary's far side through the $L_2$ point whereas no mass is
lost through the primary's far side through the $L_3$ point. The whole
system eventually becomes engulfed in a relatively warm but
low-density envelope that extends for many solar radii. Finally, the
mass transfer proceeds relatively smoothly and no shocks are observed
at the surface of the secondary.  Moreover, it is observed that the
mass transfer stream in between the two stars is relatively cooler
than the surrounding envelope, since its expansion comes at the
expense of its own internal energy.

\begin{figure}
  \begin{center}
    \includegraphics[scale=0.36,angle=-90.]{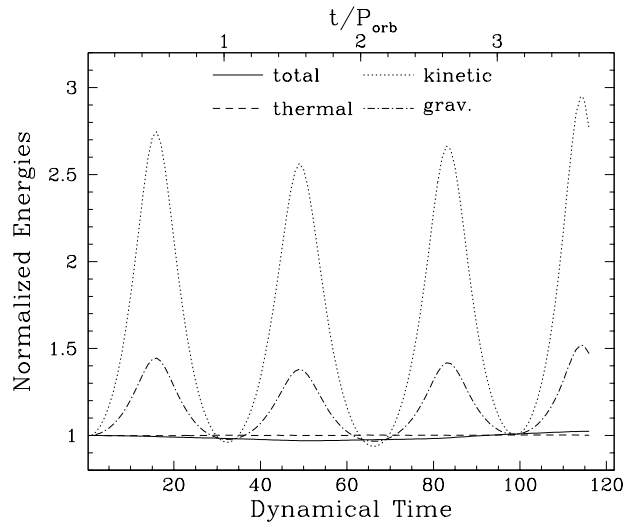}
    \caption{Different energies as a function of time for the
      0.80+0.48 M$_{\odot}$ binary with $e=0.25$.}
    \label{fig:energies_R006}
  \end{center}
\end{figure}

Figure \ref{fig:energies_R006} shows the different energies,
normalized to their initial value, as a function of time for the same
system.  The total energy is fairly well conserved during the whole
duration of the simulation. It varies by at most $\sim3\%$ and seems
to do so periodically. The eccentricity of this system is obvious from
the shape of the curve of the kinetic energy as it peaks at
periastron, halfway through the orbital period, and decreases almost
to its initial value. The different values of the extrema of the
kinetic energy suggest that the orbital separation is
changing. Similar behaviours are also observed for models with
different eccentricities.  As for the gravitational energy, it varies
in the same way as the kinetic energy, whereas the thermal energy
stays constant to within less than $0.5\%$ over the whole duration of
the simulation.
The total angular momentum, on the other hand, varies around its
initial value, by no more than 2.5$\%$ over the whole duration of the
simulation. We use the binary's centre of mass as the rotation axis to
calculate the total angular momentum of the system and all of the
angular momentum is, as expected, in the $z$-direction, i.e.\
perpendicular to the orbital plane. The angular momentum in the other
directions is at least 4 orders of magnitude smaller and remains
negligible for the whole duration of the simulations.  These
variations of the total angular momentum observed in our simulations
are acceptable given that the angular velocity of the ghosts is
artificially maintained at a fixed value (see Paper I).

\subsection{1.50 M$_{\odot}$ + 1.40 M$_{\odot}$}
\begin{figure}
  \begin{center}
    \includegraphics[angle=-90.,scale=0.7]{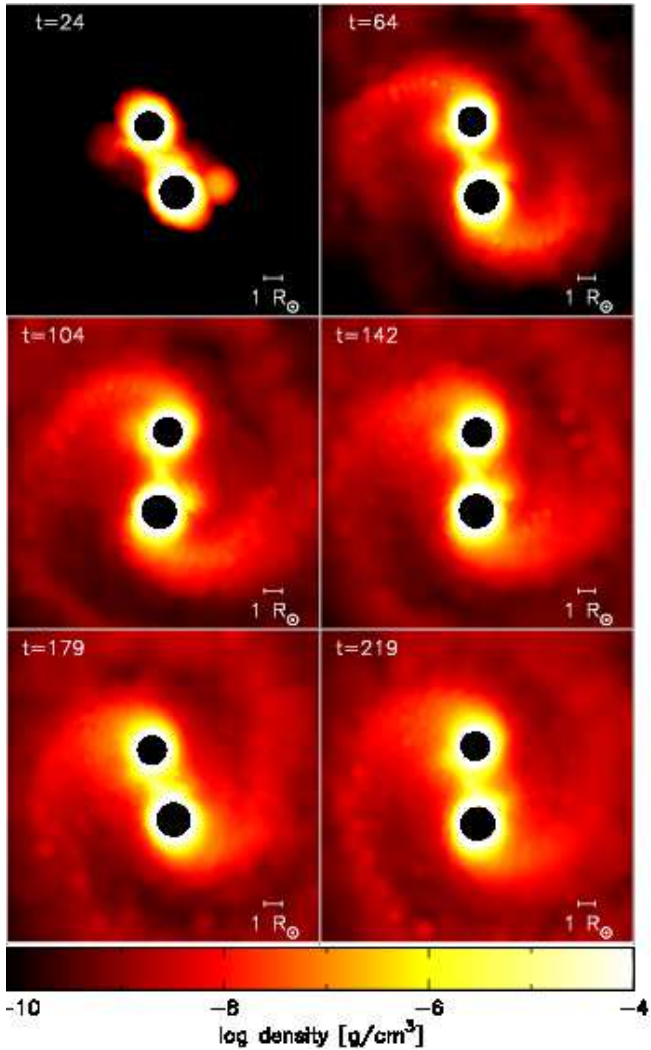}
    \caption {Logarithm of the density in the $XY$-plane for the
      $1.50+1.40$ M$_{\odot}$ binary and $e=0.25$.  Each image is 18
      R$_{\odot}$ by 18 R$_{\odot}$ and the central point masses are
      not shown.  The time is shown in units of the dynamical
      timescale ($\tau_{dyn}$) and the orbital period is $\sim 39$
      $\tau_{dyn}$.}
     \label{fig:XY_R012}
  \end{center}
\end{figure}
The second system we model is a higher-mass binary representative of
the population of relatively old open clusters. Also, since the
secondary is much larger and its density is of the same magnitude as
the primary, we expect the infalling material to interact much more
dynamically with the envelope of the secondary. The two stars are
initially set at a separation of 6 R$_{\odot}$, at apastron, which
places them well within their Roche lobe.  The location of the
boundaries is chosen at 75$\%$ of the total radius of the stars,
corresponding to a radius of $1.05$ and $0.90$ R$_{\odot}$ for the
primary and the secondary respectively.  The total number of particles
is $\sim 440,000 $ and, consequently, the total mass in SPH particles
in the primary amounts to $\sim$$1.16\times 10^{-3}$ M$_{\odot}$
whereas the secondary contains $\sim$$1.65\times 10^{-3}$ M$_{\odot}$
of SPH particles. The remainder of the mass is contained in the
central point masses.  Figure \ref{fig:XY_R012} shows the logarithm of
the density in the $XY$ plane for the $e=0.25$ case at different
times. The orbital period for this system is $\sim 39$
$\tau_{dyn}$. The interaction between the two stars is much stronger
here, as material from both stars is lost, and a clear spiral pattern
is observed and most prominent towards the end of each mass transfer
episode (i.e.\ after each periastron passage). At low eccentricity,
the mass transfer is rather smooth and has little effect on the
secondary, whereas for our largest eccentricity runs, the systems
almost come into contact at periastron and material from the primary
plows through the secondary's envelope, which is pushed around the
whole system. Most of the envelope surrounding both stars is
relatively hot as it gets heated up after the first periastron
passage. Also, unlike the low-mass binary, we observe mass loss
through both the $L_2$ and $L_3$ points, which may be enhanced by the
fact that asynchronism is substantial at periastron, thus lowering the
potential at the $L_3$ point (see $\S$\ref{sect:roche}).

\begin{figure}
  \begin{center}
    \includegraphics[scale=0.36,angle=-90.]{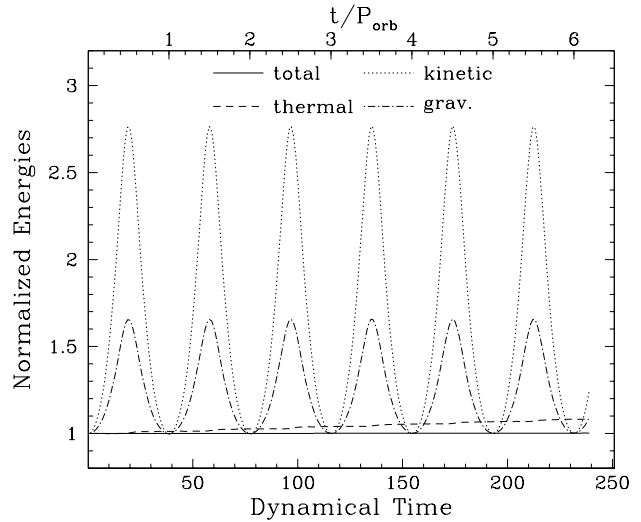}
    \caption{Different energies as a function of time for the
      1.50+1.40 M$_{\odot}$ binary with $e=0.25$.}
    \label{fig:energies_R012}
  \end{center}
\end{figure}

We show the different energies and total angular momentum for the same
simulation in Figures \ref{fig:energies_R012}. The different energies
oscillate as a function of the orbital position, with the kinetic and
gravitational energies reaching extrema at periastron.  The kinetic
energy always peaks at the same value and comes back to its initial
value when at apastron, suggesting that the binary is well relaxed and
that it follows the orbit it was initially put on. Also, the total
energy changes by no more than $\sim 0.5\%$ over the whole duration of
the simulation. We also notice that the total internal energy slowly
increases, by $8\%$ at the end of the simulation. This change in
thermal energy comes at the expense of gravitational energy, but
although $8\%$ seems substantial, we emphasize that the total thermal
energy represents roughly only 1 part in 1000 of both the kinetic and
gravitational energies.  Therefore, it would be hard to observe such a
small change in gravitational energy on the scale of Figure
\ref{fig:energies_R012}.  The total angular momentum of the system and
of the two stellar components also remains constant during the entire
simulation to a $1\%$ level for the whole duration of the simulation.

\subsection{Mass transfer rates}
\label{sect:transfer}

We now determine the mass transfer rates from our simulations. We use
the method based on the total energy of each SPH particle, as
discussed in Paper I (see also \citet{lombardi2006}), to determine to
which component SPH particles are bound. Particles are assigned to one
of the following components: the primary and secondary stars, the
binary envelope, and the ejecta.

\subsubsection{Rate and duration of mass transfer}
\begin{figure}
  \begin{center}
    \includegraphics[scale=0.425]{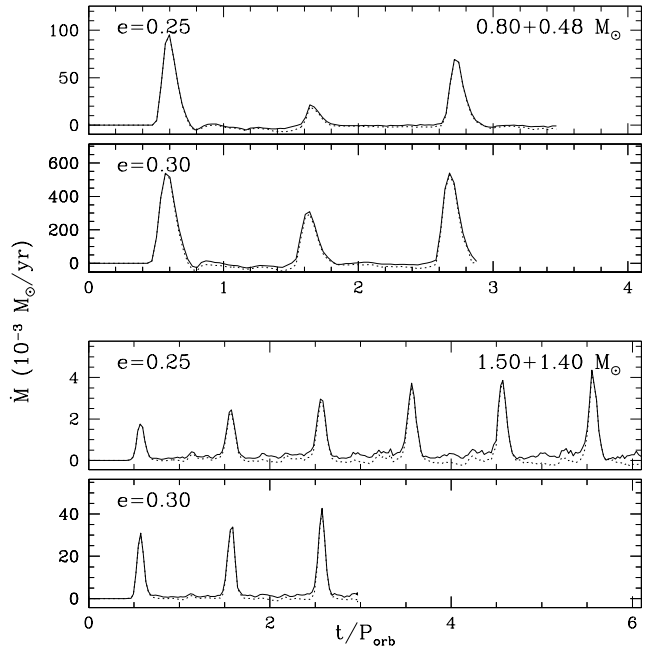}
    \caption{Mass transfer rates as a function of orbital period and
      eccentricity for a selection of runs from the $0.80+0.48$
      M$_{\odot}$ (upper two panels) and the $1.50+1.40$ M$_{\odot}$
      (lower two panels) binaries. The solid and dotted lines
      represent the mass transfer and accretion rates of the primary
      and secondary respectively.}
    \label{fig:mdots}
  \end{center}
\end{figure}
Figure \ref{fig:mdots} shows some of the mass transfer and accretion
episodes a function of time and eccentricity for the stellar
components in the $0.80+0.48$ M$_{\odot}$ and $1.50+1.40$ M$_{\odot}$
systems.  For the primary (solid lines), we plot the negative of the
mass transfer rates so that we can compare it to the (positive)
accretion rate of the secondary. In most cases, the mass transfer
rates are well defined and peak right after the periastron
passages. We note that mostly particles from the outer envelope of the
primary only are transferred during each episode and that the boundary
never becomes involved in the interaction.  For the low-mass binary,
mass transfer occurs only for eccentricity greater than $\sim 0.20$.
In these cases, the mass transferred from the primary is almost
totally accreted by the secondary, as shown by the reciprocity of the
solid and dotted lines. We note also that both rates sometime dip in
the negative part of the plots, meaning that some material is falling
back onto the primary or that the secondary is losing some of its
newly-accreted mass.  For lower eccentricities (e.g.\ $e=0.10-0.20$),
only a few particles are actually transferred and either the mass
transfer is insignificant or the code fails to correctly handle the
few tens of particles wandering in between the two stars. Finally, we
note that there is no trend in the maximum mass transfer rates of the
low-mass binary, although the need for a larger number of orbits might
be required to observe any such trends. The changes in the maximum
mass transfer rates observed are likely due to the fact that the stars
do not remain on their initial eccentric orbit, as can also be seen in
Figure \ref{fig:energies_R006}, therefore changing the degree of
overflow and the mass transfer rate.

\begin{figure}
  \begin{center}
    \includegraphics[scale=0.46]{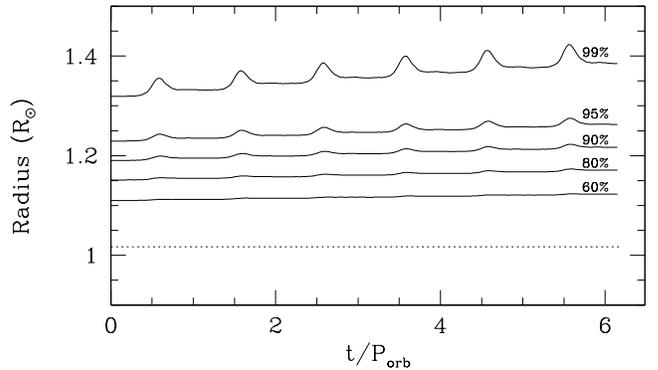}
    \caption{Radii enclosing different fractions of the total bound
      mass (in SPH particles) to the primary star as a function of
      time for the $1.40+1.50$ M$_{\odot}$ binary with $e=0.25$. The
      dotted line represents the location of the boundary.}
    \label{fig:encmassR012}
  \end{center}
\end{figure}

The high-mass binary simulations, on the other hand, all display the
characteristic episodic mass transfer peaks, with increasingly larger
mass transfer rates. Interestingly, the maximum mass transfer rates
once again all peak shortly after periastron, although the two
smallest eccentricities show quite a bit of noise in between these
peaks.  The noise is caused by material falling back onto either or
both stellar components in between periastron passages. In most
instances, the material lost by the primary is almost all accreted
onto the secondary, although the accretion rates of the secondary
shows some differences with respect to the mass loss rates of the
primary, suggesting that some mass is lost from the system.  The
$e=0.20$ case is rather noisy and there seems to be a significant
fraction of the mass transferred that falls back onto the primary and
secondary after the main episodes of mass transfer. This seems to be
important only in the smaller eccentricity cases. The primary's mass
transfer rates rarely becomes negative, unlike the secondary's
accretion rates, which are mostly negative in between periastron
passages, suggesting that the secondary loses mass. We note however
that mass becomes bound to the secondary (and the primary) only
temporarily as subsequent episodes of mass transfer are sometimes
energetic enough to plow through the surrounding envelope of the
secondary and eject some of this material. Likewise, the maximum mass
transfer rates are observed to increase both with time and
eccentricity. Given that both stars remain very close to their initial
eccentric orbit (see Figure \ref{fig:energies_R012}), this increase in
the peak mass transfer rate is likely due to an increase of the
primary star radius.  Figure \ref{fig:encmassR012} shows the radii
enclosing different fractions of the total bound mass for the
primary. The periastron passages are clearly visible and, most
importantly, the radii \textit{in between} the mass transfer episode
gradually increase, which is indicative of the expansion of the
primary's envelope as matter is being lost. This increase in radius
inevitably leads to an increase in the degree of overflow and,
consequently, the mass transfer rate. We also note that the boundary
is well within the star and the radius containing $60\%$ of the mass
in SPH particles barely changes with time, indicative that tidal
effects are negligible at this location.
Finally, we note the similarities between the two systems modeled in
the position, duration, and shape of the mass transfer rate
episodes. In particular, their shape is suggestive of a Gaussian
function.

The range of mass transfer rates observed in our simulations ranges
from a few $10^{-6}$ M$_{\odot}$ yr$^{-1}$, for the high-mass binary
with $e=0.15$, to $0.1$ M$_{\odot}$ yr$^{-1}$ for the low-mass
binaries. We emphasize that these relatively high mass transfer rates
last only for a short period of time (i.e. $\sim$0.20 orbit) and the
total mass transferred amounts to less than $\sim 10^{-4}-10^{-5}$
M$_{\odot}$ yr$^{-1}$ per periastron passage. In all cases, the number
of particles transferred ranges from a few hundreds to many thousands
per mass transfer episode. Binaries where the number of particles
transferred is less than $\sim100$ are considered as not transferring
mass on the basis of the poor SPH treatment for such low numbers (see
Table \ref{tab:models}). Given the least massive particles in our
simulations, the lowest possible mass transfer we can model
(notwithstanding the numerical noise), is of the order of
$10^{-7}-10^{-6}$ M$_{\odot}$ yr$^{-1}$, which is comparable to that
of \citet{d'souza2006}. However, given the low number of particles
that would be transferred in such instances, the SPH approach fails at
properly evaluating the hydrodynamical forces on these isolated
particles. We compare our results with theoretical expectations in
$\S$\ref{sect:comparisons}.

\subsubsection{Gaussian fits to mass transfer episodes}
\label{sub:gauss}
\begin{figure}
  \begin{center}
    \includegraphics[scale=0.42]{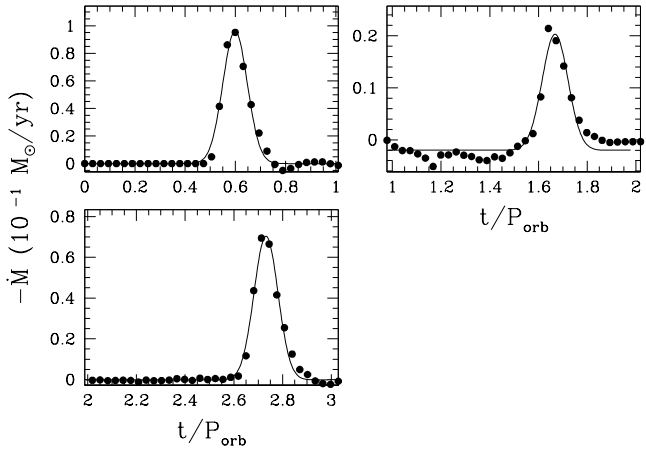}
    \caption{Gaussian fits to the primary's mass transfer episodes for
      the 0.80+0.48 M$_{\odot}$ binary with $e=0.25$. The values of
      the fitted parameters are reported in Table
      \ref{tab:gaussfits}.}
    \label{fig:gaussfit2R006}
  \end{center}
\end{figure}
\begin{figure}
  \begin{center}
    \includegraphics[scale=0.42]{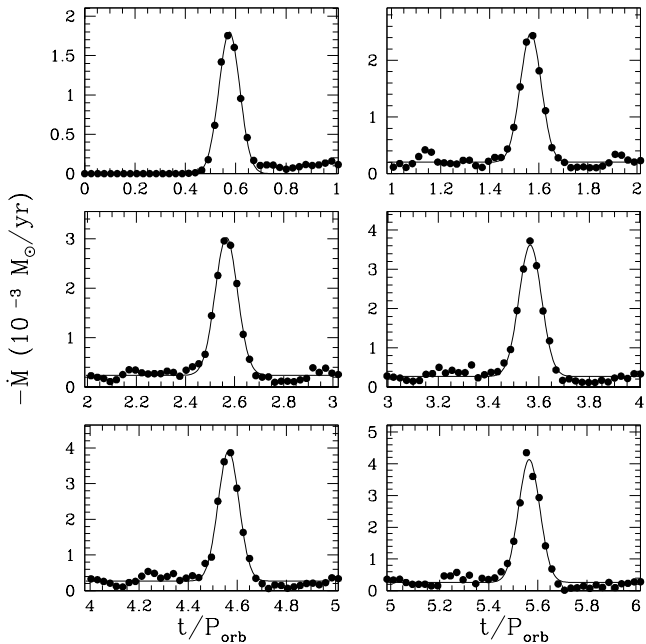}
    \caption{Gaussian fits to the primary's mass transfer episodes for
      the 1.50+1.40 M$_{\odot}$ binary with $e=0.25$. The values of
      the fitted parameters are reported in Table
      \ref{tab:gaussfits}.}
    \label{fig:gaussfit2R012}
  \end{center}
\end{figure}
Using our mass transfer rate profiles of Figure \ref{fig:mdots}, we
now fit a Gaussian function to every mass transfer episode. The
general Gaussian we use has the following form:
\begin{equation}
  \dot{M}(t) = A \exp\Big(-\frac{(t-\mu)^2}{2\sigma^2}\Big) + D
  \label{eq:gaussian}
\end{equation}
where $A$ is the maximum amplitude, $\mu$ is the centre of the
Gaussian, $\sigma$ is proportional to the width of the Gaussian and
$D$ is the background (or continuum) mass transfer rate. The latter
parameter is used to measure the background noise as in some case the
mass transfer rates do not fall back to zero in between mass transfer
episodes. All the free parameters are fitted using the nonlinear
least-squares method of Levenberg-Marquardt \citep{press1992}.  We fit
the height of the Gaussian extended wings so that the width of the
Gaussian matches more closely the data points. However, in cases where
matter falls back onto the stars between periastron passages, the
fitting procedure is to be taken with care. Examples of the Gaussian
fits to the mass transfer episodes are shown in Figures
\ref{fig:gaussfit2R006} and \ref{fig:gaussfit2R012} for two of our
simulations. The parameters obtained from the fitting procedure are
given in Table \ref{tab:gaussfits}. Some data points are assigned a
relatively large error since they are part of the pre-periastron mass
transfer episodes and do not contribute to the main episode of mass
transfer nor to the fitting procedure. Note that doing so does not
significantly change the values of both $\mu$ and $\sigma$.  Moreover,
for every first episode of mass transfer, we do not fit the continuum
(parameter $D$) as we expect the value of the mass transfer rate prior
to the first periastron passage to be identically zero. We do fit this
parameter for any subsequent peak however.

For most of our simulations, Gaussians fit the data points remarkably
well.  In most cases, the amplitude, centre, and width all closely
match the data points. Again, for cases where matter is observed to
fall back onto the stars, the fits to the height of the extended wings
is obviously not as reliable. For example, the Gaussian fit for the
low-mass binary with an eccentricity $e=0.20$ is not reliable as the
mass transfer is too noisy.  An important source of uncertainty on
these fitted parameters, especially $A$ and $D$, is the noise on
either side of the peaks seen in the data.  On the other hand, the
width and amplitude of most (if not all) of the mass transfer episodes
are well matched by Gaussians.

\begin{figure}
  \begin{center}
    \includegraphics[scale=0.42]{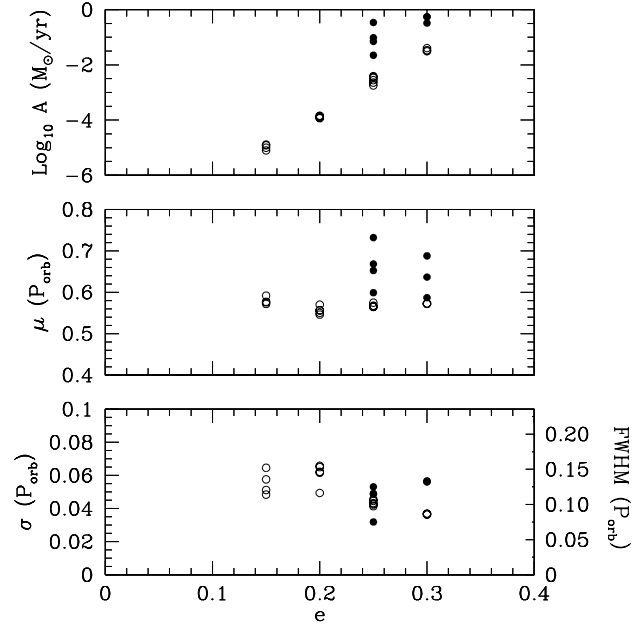}
    \caption{Amplitude, position, and width of the Gaussian fits to
      the mass transfer episode for the primary as a function of
      eccentricity for both binary systems. Solid dots are for the
      $0.80+0.48$ M$_{\odot}$ binary while open dots are for the
      $1.50+1.40$ M$_{\odot}$ binary. See also Table
      \ref{tab:gaussfits}.}
    \label{fig:trends1}
  \end{center}
\end{figure}

We plot, in Figure \ref{fig:trends1}, the amplitude, centre, and width
of all the Gaussians we fitted as a function of eccentricity. Many
trends can be seen in this plot. First, the maximum mass transfer rate
increases with the eccentricity.  This is expected since as the
eccentricity increases, the periastron distance gets smaller and the
two stars get closer to each other, thus facilitating mass
transfer. Our results also suggest that the maximum mass transfer rate
increases linearly with the eccentricity, although we also expect a
cut-off at low eccentricity where the primary will not fill its Roche
lobe even when at periastron. Also, although we only have two data
points from our low-mass binary simulations, these two simulations
suggest a similar trend.  As for the position where the maximum
mass transfer rate occurs, the results from our high-mass binary
simulations clearly show that mass transfer rates peak at an orbital
phase slightly larger than periastron, around $0.55-0.57$.
\noindent 
Although mass transfer starts around periastron, it only peaks later
when the two stars have already started getting further away from each
other. This is in contrast with one of the basic assumptions of SWKR07
and SWKR09, who assumed that mass transfer occurred instantaneously at
periastron.  The observed delays in the peak mass transfer rates are
consistent with a free-fall time ($\tau_{ff}\simeq
0.5(G\bar{\rho})^{-1/2}$) into the secondaries' potential well at
periastron.  Following \citet{eggleton1983}, the free fall times for
our systems should be about $0.05-0.065$ P$_{orb}$, similar to the
delays observed in our simulation (see e.g.\ Table 2), with an
expected slight increase for the low-mass binaries. It seems likely
therefore that the small differences in the position of the mass
transfer rate peaks observed in our simulations are real. Our method
for determining the mass transfer rates does not tell whether a
particle \textit{will} be transferred but rather if it has been
transferred, which is what we define as mass transfer, and our results
seem to indicate that this occurs over a free-fall time.  Only at
periastron is the tidal force large enough to strip the deeper layers
of the primary. Since this material has to travel to the Roche surface
before being assigned to the secondary, a delay in the peak mass
transfer rate is to be expected. We note that such delays in the peak
mass transfer rates should be intrinsic to eccentric binaries as these
systems never exactly fill their Roche lobe but rather periodically
shrink within and expand beyond it.  Although the results from our
low-mass binary simulations are less suggestive, the delays observed
in the position of the maximum mass transfer rate also suggest that
the maximum degree of overflow should occur later than periastron.

Finally, we also observe in both sets of simulations that the width
(or duration) of the mass transfer episode is finite in time and
arguably independent of the eccentricity. Our results suggest that the
full width at half maximum (FWHM $\approx \sqrt{8\ln 2}\sigma$) is
approximately $0.10-0.13$ P$_{\textrm{orb}}$.  One could argue that
there is a small negative slope suggesting that the higher the
eccentricity, the faster the mass transfer occurs, which is plausible
since stars on high eccentric orbits spend less time around periastron
compared to star on low eccentric orbits.  No matter the trend, this
value of the width of the mass transfer rates is also in contrast with
another basic assumption used by SWKR07 and SWKR09, who assumed an
instantaneous mass transfer rate. Our results clearly show that this
is not the case and that the mass transfer occurs over an extended but
finite period of time.

\subsection{Accretion onto the secondary}
\label{sect:accretion}
\begin{figure}
  \begin{center}
    \includegraphics[scale=0.6]{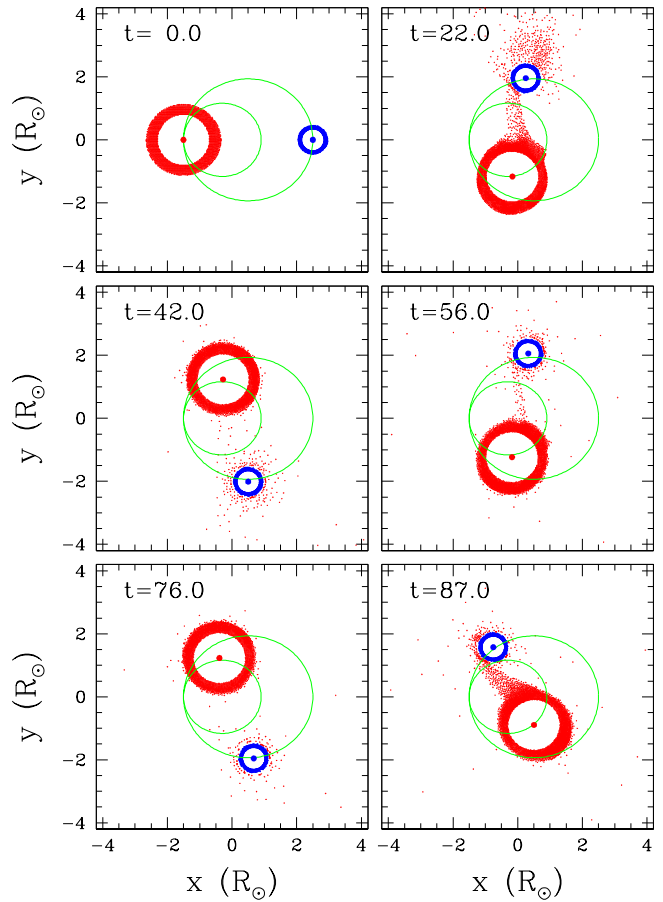}
    \caption{Origin of particles in the orbital plane for the
      0.80+0.48 M$_{\odot}$ binary with $e=0.25$. Red and blue dots
      are particles that initially come from the primary and secondary
      respectively.  The time is shown in units of the dynamical
      timescale ($\tau_{dyn}$) and the orbital period is $\sim 32$
      $\tau_{dyn}$.In this case, the secondary is not affected by the
      infalling material.}
    \label{fig:originR006}
  \end{center}
\end{figure}
\begin{figure}
  \begin{center}
    \includegraphics[scale=0.47]{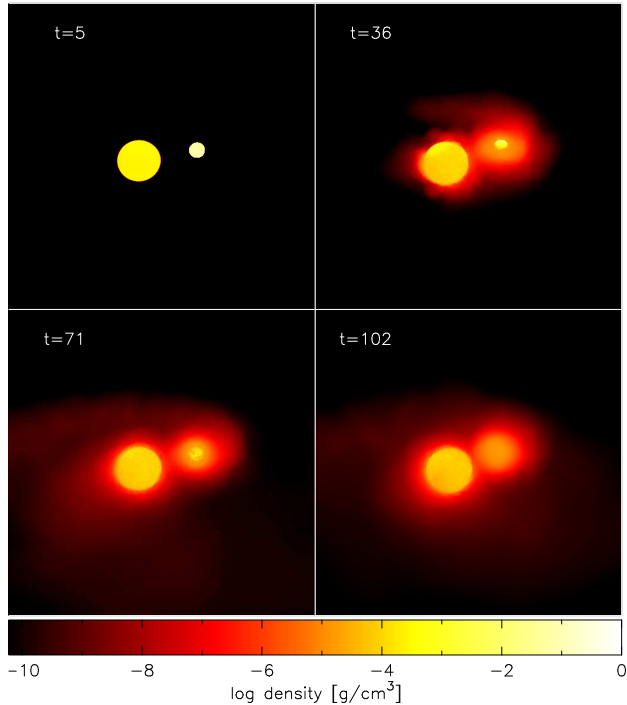}
    \caption{Surface rendition (with $\tau=0.08$) of the density for
      the 0.80+0.48 M$_{\odot}$ binary with $e=0.25$ showing the
      accretion around the secondary.  The time is shown in units of
      the dynamical timescale ($\tau_{dyn}$) and the orbital period is
      $\sim 32$ $\tau_{dyn}$. Initially, a thick disk forms around the
      secondary, but later engulfs it almost completely. Mass loss
      occurs also primarily from the secondary's far side, at the
      $L_2$ point.}
    \label{fig:3D_R006}
  \end{center}
\end{figure}
Most of the mass lost by the primary eventually becomes bound to the
secondary.  One way to look at the accretion is to look at the origin
of the particles making up each component. This is shown in Figure
\ref{fig:originR006}, where we plot the origin of the particles in the
orbital plane and where red and blue dots are particles that were
initially bound to the primary and secondary respectively. This
colour-coded representation allows us to track the particles as they
are shuffled around and become bound to any of the components of the
system (i.e.\ the secondary, the binary envelope, or the ejecta). In
the case of the low-mass binary, we see that the secondary is not
strongly affected by the infalling material as none of its own
particles are being mixed up with the material from the primary.  As a
matter of fact, the secondary is so dense that it is not perturbed at
all by the mass transfer episodes and its accreted material simply
forms an envelope around it.  Despite this large density gradient at
the surface, we note that the interaction between the infalling
material from the primary and the secondary's envelope is relatively
smooth and no shocks are observed at the surface of the boundary.  The
smoothing lengths of the transferred particles are consistently
adjusted in time (see $\S$\ref{sect:method}) and are of the same order
of magnitude as those of the particles at the surface of the
secondary. Also, the difference in mass between the particles being
transferred and the particles forming the outer layers of the
secondary differs by less than one order of magnitude and we do not
observe spurious motions in the envelope suggesting interactions
between particles with extreme mass ratios (see e.g.\
\citealt{lombardi1999}).  The density contrast can also be observed in
Figure \ref{fig:3D_R006}, which shows a \verb|SPLASH|
three-dimensional surface rendition of the $0.80+0.48$ M$_{\odot}$
system with $e=0.25$. The surfaces are rendered by defining a critical
surface through which we can not see, similar to an optical depth.
This three-dimensional rendition allows for the visualization of the
whole system. We see that the material transferred from the primary
initially forms a thick torus-like cloud around the secondary, and
subsequent episodes of mass transfer eventually form an envelope that
rather engulfs the secondary and corotates with it.  Moreover, the
primary also becomes engulfed by a thin envelope. Since the secondary
does not lose any mass, this envelope, as well as the binary ejecta,
is made up of the material from the primary.
\begin{figure}
  \begin{center}
    \includegraphics[scale=0.6]{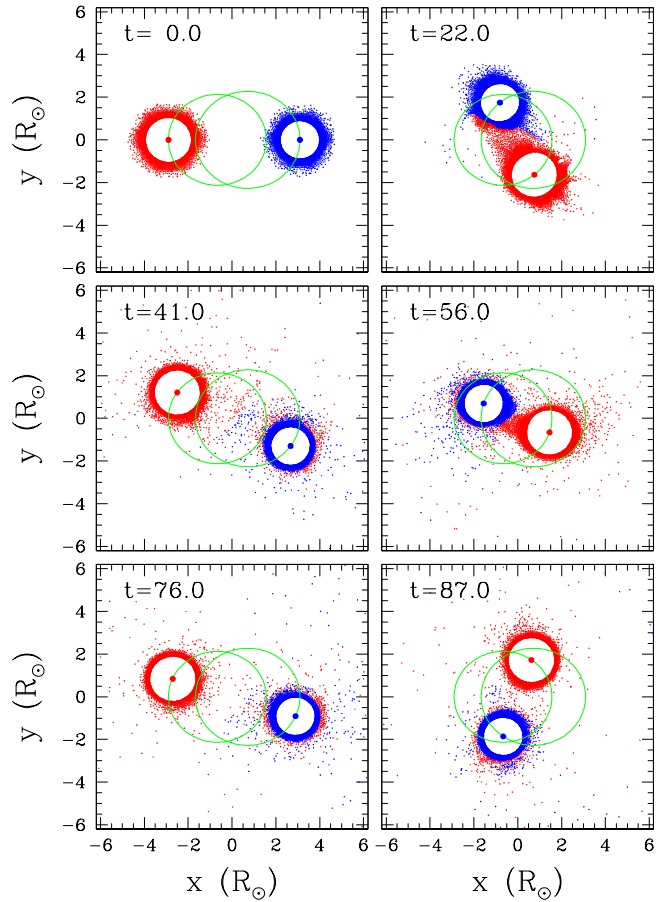}
    \caption{Origin of particles in the orbital plane for the
      1.50+1.40 M$_{\odot}$ binary with $e=0.30$.  Red and blue dots
      are particles that initially come from the primary and secondary
      respectively.  The time is shown in units of the dynamical
      timescale ($\tau_{dyn}$) and the orbital period is $\sim 39$
      $\tau_{dyn}$.  In this case, the secondary loses material because
      of partial Roche lobe overflow and the interaction of the
      infalling material with its envelope.}
    \label{fig:originR013}
  \end{center}
\end{figure}
\begin{figure}
  \begin{center}
    \includegraphics[scale=0.47]{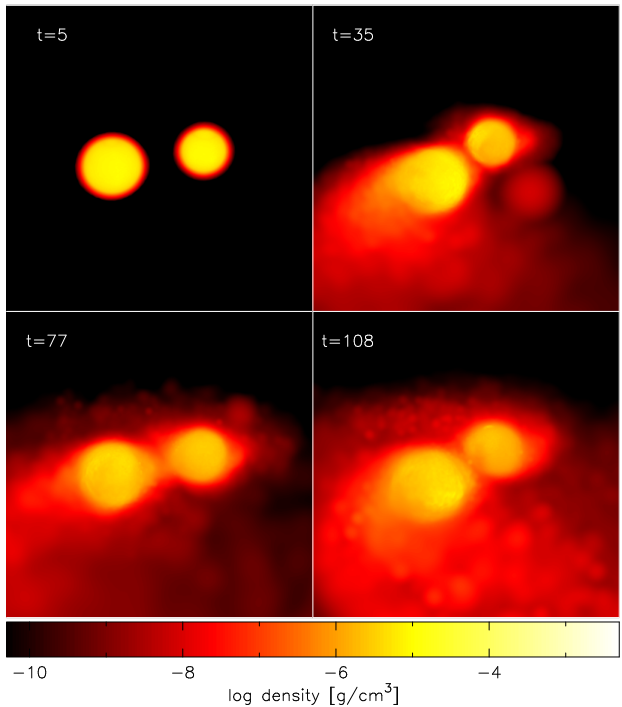}
    \caption{Surface rendition (with $\tau=25.0$) of the density for
      the 1.50+1.40 M$_{\odot}$ binary with $e=0.30$ showing the
      accretion around the secondary.  The time is shown in units of
      the dynamical timescale ($\tau_{dyn}$) and the orbital period is
      $\sim 39$ $\tau_{dyn}$.  In this case, an envelope engulfs both
      star and mass loss occurs also from the far side of both stars,
      at the $L_2$ and $L_3$ points}
    \label{fig:3D_R013}
  \end{center}
\end{figure}

In the case of the high-mass binary, the similar densities of the two
stars allows for the material being transferred to interact much more
strongly with the secondary's envelope, as shown in Figure
\ref{fig:originR013} (for $e=0.30$).  As shown by the red and blue
dots, particles from the outermost layers of the secondary, as well as
from the primary, are lost from both stars and scattered everywhere in
the orbital plane.  Although the secondary loses almost as many
particles than the primary, the total mass lost by the secondary
actually remains relatively small ($\sim5\%$ for the $e=0.30$ and
$\sim 10\%$ for the $e=0.25$ case, at the end of both simulations).
Again, we emphasize that the fact that the secondary loses some mass
is the result of both the secondary slightly overfilling its Roche
lobe (see upper right panel of Figure \ref{fig:originR013} for
example) and the interaction of the infalling material plowing through
the envelope of the secondary.  This is more analogous to a so-called
``direct impact'' where the secondary fills most of its Roche lobe and
therefore is hit almost right after the transferred material passes
the $L_1$ point. Interestingly, we do not observe such a mass loss
from the secondary for smaller eccentricities (e.g.\ $e=0.15$ and
$e=0.20$), which agrees with the fact that only at (relatively) high
eccentricity does the system come close to contact.  As can be also
seen in Figure \ref{fig:3D_R013}, all of the material being lost from
both stars eventually engulfs the whole system rather than forming an
envelope around the secondary only.  This envelope is substantially
denser and thicker than in the low-mass binary case (see Figure
\ref{fig:3D_R006}) as the surface rendition of Figure
\ref{fig:3D_R013} uses a much larger optical depth in order to peer
through the envelope and see the surface of the stars.  We observe the
formation of such an envelope in all of our simulations for this
binary, although the envelope for the $e=015-0.20$ cases is thinner as
less material is lost from either star. We note that in the high-mass
binaries, the particle mass profiles in both stars are almost
identical and we expect and, in fact, observe no spurious motion in
the envelope. Similarly, the smoothing lengths are consistently
evolved such that particles in regions of similar density have similar
smoothing length, allowing for a better spatial resolution.

As in the case of the mass transfer episodes, the accretion episodes
also display similar positions and duration and characteristic shapes
of Gaussian functions.  Our results are again displayed in Table
\ref{tab:gaussfits}.  The position of the peaks and the width of the
Gaussian are similar to the mass transfer episodes of the primary.
The Gaussian parameters for the accretion onto the secondary follow
the same trends as shown in Figure \ref{fig:trends1}, with only minor
difference in the width of the Gaussians ($\sigma$). Indeed, we notice
only a slightly larger spread in $\sigma$ for the accretion rate when
compared with the mass transfer rates.  Although we could expect the
maximum mass transfer rate ($A_1$) to be larger than the maximum
accretion rate ($A_2$), we see that it is not always the case.  The
reason is simply that the values of the continuum (parameter $D$) are
different for the primary and the secondary, thus yielding a different
zero level from which the peak values are measured. In any case
however, the total mass accreted by the secondary is always less (or
equal) to the mass lost by the primary. The remaining mass is
obviously lost to the binary as a whole or to the ejecta, which is
what we discuss next.

\subsection{Mass loss}
\label{sect:loss}
We now quantify the amount of mass lost to either the binary envelope
or the ejecta in our simulations. Particles are assigned to either of
these two components if they are far enough from either stellar
components and/or their total (relative) energy is positive (see Paper
I).

\subsubsection{Escaping Particles}
Escaping particles are particles that are found far (i.e.\ many
smoothing lengths) from the bulk of the particles. By design, the
code, and more specifically the tree building and the neighbours
search, run into some problems when particles escape and/or are found
in between the two stars, with only a few other close
particles. Indeed, when particles are ejected from the system, the
search for the required number of neighbours become lengthy and
sometimes unsuccessful. To circumvent this issue, we set a maximum
smoothing length ($\sim$5 R$_{\odot}$) such that the code does not
unnecessarily spend CPU time iterating and adjusting the smoothing
length of a small set of escaping particles (see also
$\S$\ref{sect:method}).  These escaping particles lack sufficient
neighbours, but their low numbers and total mass are small and do not
affect the (hydro)dynamics of the mass transfer process
itself. 

\subsubsection{Binary envelope and ejecta}
\label{sub:envel}
We now assess the degree of mass loss during the mass transfer
episodes in our simulations. We find that in all cases, the mass
contained in the binary envelope is greater than that in the ejecta,
by a factor of at least two.  Both the binary envelope and the ejecta
grow in mass as a function of time, and although no clear episodes of
mass growth is observed for the ejecta, we observe a stepwise increase
in the mass bound to the binary envelope.  The total mass in each
component ranges from a few $10^{-7}$ M$_{\odot}$ for the high-mass
binary to $\sim6 \times 10^{-5}$ M$_{\odot}$ for the low-mass binary.
This amounts to $\sim 0.1-1 \%$ of the total initial mass (in SPH
particles) of the primary.

\begin{figure}
  \begin{center}
    \includegraphics[scale=0.43]{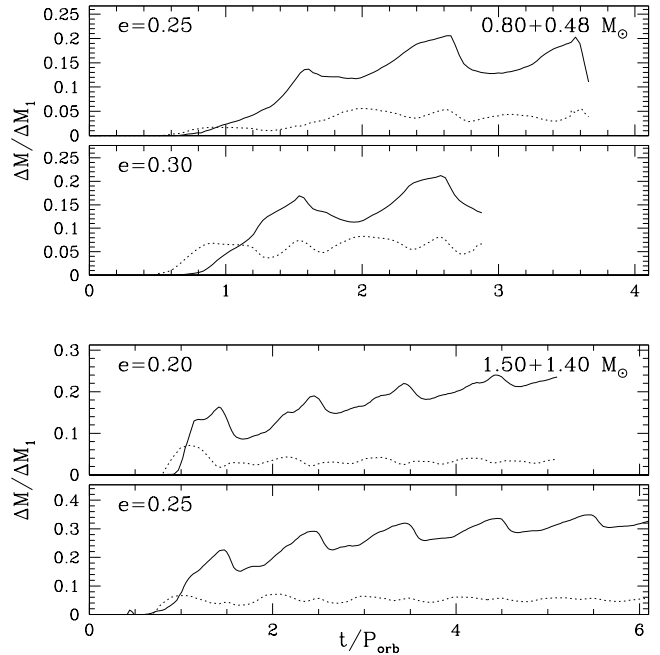}
    \caption{Changes in mass in the binary envelope and ejecta
      normalized by the change in mass of the primary for selected
      runs for both the 0.80+0.48 M$_{\odot}$ (upper two panels) and
      the 1.50+1.40 M$_{\odot}$ (lower four panels) binaries. In all
      cases, $\sim5\%$ of the mass lost by the primary ends up in the
      ejecta.}
    \label{fig:pctejected}
  \end{center}
\end{figure}

Figure \ref{fig:pctejected} shows the change in bound mass of the
binary envelope and the ejecta normalized by the change in mass of the
primary. Essentially, this shows the fraction of the mass lost by the
primary that ends up in the binary envelope or the ejecta.
Interestingly, the fraction of mass bound to the binary envelope shows
some periodic behaviour, peaking shortly \textit{before} periastron,
where the stars are at their closest separation along the orbit.
This lag comes from the fact that (some of) the infalling material
only temporarily becomes bound to the secondary before becoming bound
to the envelope or the ejecta. By the end of our simulations, around
$20 \%$ of the mass lost by the primary end up in the binary envelope,
and our results suggest that this fraction slowly increases as a
function of time.

The mass in the ejecta, on the other hand, is roughly constant around
$5\%$ for all of our simulations.  This is an unexpected result, given
that we observe this trend in all of our simulations, no matter the
mass of the stars or the eccentricity. Moreover, given that the degree
of mass loss in binary evolution is rather unconstrained (e.g.\
SWKR09), this result is suggestive of an almost uniform mass loss over
different binary masses.  Although conservative mass transfer is
usually employed as an idealized case, the constant and small fraction
of mass lost in our simulations suggests that mass transfer is indeed
non-conservative but only to a small degree.  This is also interesting
in the context of two (slightly) different accretion scenarios, where
accretion occurs via an accretion disk or through a direct impact. In
the latter case, when the secondary fills a significant fraction of
its Roche lobe, the matter falling in from the $L_1$ point hits the
accretor almost directly, whereas in the former case, material falls
deep in the potential well of the secondary and forms a disk.
Although the direct impact scenario is more representative of our
high-mass binary rather than the low-mass binary, and whether or not
the infalling material interacts with the secondary's envelope, we
still get that roughly $5\%$ of the mass lost by the primary ends up
in the ejecta.  Note, however, that our simulations do not allow us to
assess the fate of the binary envelope. i.e.\ whether it is going to
be expelled from the system and become part of the ejecta or be
accreted by either star.

\section{Comparisons with previous work}
\label{sect:comparisons}
Similarly to \citet{church2009}, our results show that the mass
transfer rates get increasingly larger as the stars get closer to each
other at periastron. Moreover, our results indicate that the mass
transfer episodes do not occur precisely at periastron and last for a
constant fraction of the orbital period, independent of the
eccentricity.  However, unlike \citet{church2009}, our mass resolution
does not allow us to resolve mass transfer rates as low as
$\sim10^{-10}-10^{-9}$ M$_{\odot}$ yr$^{-1}$.  Having a better mass
resolution would help increase the number of particles in the stream
of material for our low-eccentricity binaries, but we do not expect
that this would drastically change our conclusions.

Our binaries are set up such that they are in corotation at apastron,
therefore making them subsynchronous at periastron. In both of our
sets of simulations, the ratio of the angular velocity, which is fixed
for the whole duration of the simulations, to the orbital velocity at
periastron ranges from around $0.30$ to $0.60$.  Thus, according to
\citet{sepinsky2007a}, this has the effect of slightly increasing the
Roche lobe radius, by $\sim 5\%$, when compared to the instantaneous
Roche lobe radius at periastron.  Such an increase would therefore
effectively \textit{decrease} the degree of overflow and,
consequently, the mass transfer rate. 

\begin{figure}
  \begin{center}
    \includegraphics[scale=0.46]{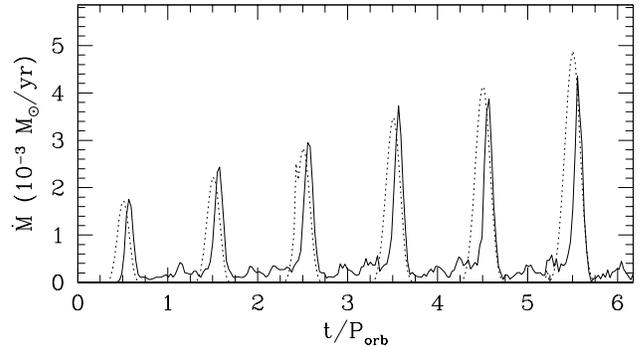}
    \caption{Comparison between the instantaneous mass transfer rate
      of Equation \ref{eq:mdot_gokhale} (dotted) and our results
      (solid) for the the $1.50+1.40$ M$_{\odot}$ binary with
      $e=0.25$.}
    \label{fig:mdotscompare}
  \end{center}
\end{figure}

Comparing the magnitude of the mass transfer rate observed in our
simulations with theoretical expectations is difficult because
estimates of the actual radius of the primary from our simulations are
uncertain. Nevertheless, we build a simple model for mass transfer
using the (dynamical) mass transfer rate derived for polytropes of
index $n$ by \citet{pacz1972} (see also
\citealt{edwards1987,eggletonbook,gokhale2007}),
\begin{equation}
        \dot{M_1} = -\dot{M}_0 \Big( \frac{R-R_L^{inst}}{R} \Big)^{n+3/2},
        \label{eq:mdot_gokhale}
\end{equation}
where $\dot{M}_0$ is a canonical mass transfer rate. R$_L^{inst}$ is
the instantaneous Roche lobe radius, i.e.\ a simple generalization of
the Roche lobe radius for circular and synchronous binaries
\citep{eggleton1983}:
\begin{equation}
  R_L^{inst} = D(t) \frac{0.49 q_1^{2/3}}{0.6q_1^{2/3} +
    \textrm{ln}(1+q_1^{1/3})},
  \label{eq:R_Linst}
\end{equation}
where $D(t)$ is the instantaneous separation of the two stars.  This
mass transfer rate, which applies when the donor can be approximated
by a polytrope, depends strongly on the degree of overflow, as
expected, and is equally zero when $\Delta R \leq 0$ (this rate is
somewhat different than that of \citet{ritter1988}, which applies when
the photosphere is resolved).  Starting with the primary's radius as
measured at the start of our simulation, i.e. $R_1\simeq 1.8
R_{\odot}$, we calculate the instantaneous degree of overflow, $\Delta
R$, based on the instantaneous separation of the stars, and a mass
transfer rate.  This is shown in Figure \ref{fig:mdotscompare} for the
high-mass binary with $e=0.25$. We have arbitrarily set the canonical
mass transfer rate ($\dot{M_0}\simeq 10^{-1} M_{\odot}$ yr$^{-1}$)
such that the first peak of mass transfer matches that from our
simulation.  Moreover, to mimic the slightly increasing peak mass
transfer rate, as can be observed in our simulations, we assume that
the radius of the star increases at apastron passage by increments of
$0.05$ R$_{\odot}$.  We do not expect the radius of the primary to
change by much over the course of our simulations since the total mass
transferred is small, and this artificial (and rather large) change in
radius simply allows for a better match with the increasing peak mass
transfer rates.  We also note that we see no reasons why both the
canonical mass transfer rate $\dot{M_0}$ and the polytropic index $n$
should remain constant as mass transfer proceeds, although we have
assumed so here, which could compensate for changes in radius.
The difficulty in doing such a comparison \textit{with eccentric
  orbits} lies in the facts that we are using the instantaneous Roche
lobe radius derived for circular and synchronous binaries and that the
theoretical mass transfer rate used here was derived for polytropes of
constant $n$.  Although this simple model agrees qualitatively well
with our simulations (e.g.\ Gaussian-like episodes of mass transfer),
we emphasize that we arbitrarily fixed the canonical mass transfer
rate so that the peaks match.  However, we observe that the position
and width of the mass transfer episode strikingly differ from the
theoretical expectation.  The width of the peaks depends on the star's
radius, as a shrinking star would delay the start of mass transfer (as
well as decrease the degree of overflow), therefore decreasing the
mass transfer rate.  On the other hand, there is no parameter that
could account for the position of the maximum rates as, by
construction, the largest degree of overflow occurs when the stars are
closest to each other, i.e.\ at periastron.

\section{Conclusions and Future Work}
The evolution of binary stars has grown into an intense field of study
since it has become clear that many populations of stars have to form
through interactions with close stellar companions. Although the main
phases of binary evolution are nowadays well understood, these
evolutionary paths usually rely on the (idealized) formalism derived
for \textit{circular} and \textit{synchronized} orbits. This so-called
Roche lobe formalism does not apply for close and interacting
\textit{eccentric} binaries, in which the rotation is asynchronous and
the gravitational potential time-dependent.  Given the relatively
large number of binary stars, and in particular, of binary stars with
eccentric orbits, it is imperative to better understand the
interactions of these systems in order to further constrain the
different galactic populations of exotic stars.  Recent breakthroughs
by \citet{sepinsky2007a}, SWKR07 and SWKR09, in particular, have
allowed to extend the knowledge of the long-term evolution of
eccentric binaries.  Although these works clearly show that eccentric
binaries behave differently from circular ones, their conclusions are
based on a number of assumptions. In this paper, we have presented the
results of SPH simulations with the aim of constraining these
assumptions.

\begin{figure}[!t]
  \begin{center}
    \includegraphics[scale=0.53]{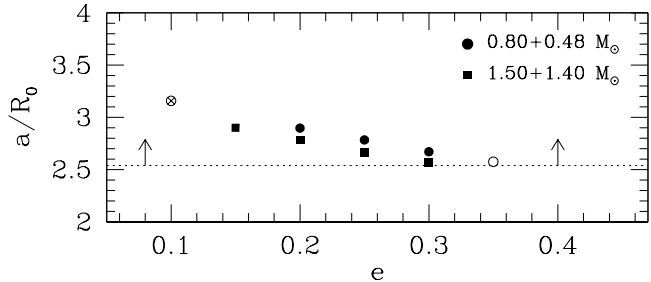}
    \caption{Semi-major axis normalized by the initial radius of the
      primary as a function of eccentricity for all of our
      simulations. Solid dots are our successful runs, open dots are
      runs for which mass transfer was too large for our boundary
      conditions to handle, and open-crossed symbols are for cases
      when mass transfer was too low or not resolved. The dotted line
      represents an approximate delimitation above which our boundary
      conditions can be applied.}
    \label{fig:space}
  \end{center}
\end{figure}
The results from our large-scale simulations are interesting both for
the performances of our alternate approach (see Paper I) and for the
characterization of the mass transfer episodes. Our boundary
conditions can effectively handle intermediate mass transfer rates
($\sim 10^{-6}-10^{-4}$ M$_{\odot}$ yr$^{-1}$), although particles
penetrate the boundary when the periastron distance is such that the
mass transfer rates become too large. On the other hand, our code, by
design, does not handle cases where only a handful of particles are
transferred.  The parameter space where our technique can be applied
is therefore restricted by these two conditions on the number of
particles.  Figure \ref{fig:space} summarizes the orbital parameters
for which our technique is well suited. We see that when the
normalized semi-major axis is greater than $\sim2.5$ R$_{\odot}$, our
boundary conditions behave well as the mass transfer rates are not
excessively large. Also, the use of more lower-mass particles would
help resolve lower mass transfer rates and allow for the modeling of
systems with higher $a/R_{\odot}$ values.

The results from our simulations of mass transfer also show clear
trends. In particular, we show that the episodes of mass transfer can
be described by Gaussians with a FWMH of $\sim 0.12-0.15$
P$_{\textrm{orb}}$, and the peak mass transfer rates occur after
periastron, around an orbital phase of $\sim 0.55-0.56$. It is
interesting to note that these results apply for both of the binary
systems modeled and for any eccentricity. The technique used in this
work represents an interesting alternative to previous work (e.g.\
\citealt{edwards1987}, \citealt{regos2005}, \citealt{church2009}) and
we suggest that our results on the properties of interacting eccentric
main-sequence binaries could be used in analytical work such as that
of SWKR07 and SWKR09 to further constrain the evolution of such
stars. We also discussed the accretion onto the secondary and showed
that it is also well characterized by similar Gaussians. The accreted
material is observed to form a rather sparse envelope around the
secondary, in the low-mass binary, and around both stars, in the
high-mass binary.  Although the fate of this envelope is not
determined using our method (whether it is going to be accreted onto
either stars or ejected from the system), we showed that a constant
fraction of the material lost by the primary is ejected from the
systems. Although poorly constrained, the concept of non-conservative
mass transfer is generally accepted nowadays and our results may help
constrain the degree of mass conservation in binary evolution. In the
future, we hope to cover more of the parameter space ($q$, $a$, and
$e$) in order to get a better picture of mass transfer in eccentric
binaries.

\acknowledgments We wish to thank the anonymous referee as well as
Doug Welch and James Wadsley for useful discussions about this
project. This work was supported by the Natural Sciences and
Engineering Research Council of Canada (NSERC) and the Ontario
Graduate Scholarship (OGS) programs, and made possible in part by the
facilities of the Shared Hierarchical Academic Research Computing
Network (SHARCNET: www.sharcnet.ca).

\clearpage
\input{tab1}
\input{tab2}

\end{document}

%% file: tab1.tex
\begin{deluxetable}{c|cccccccc}
\tabletypesize{\scriptsize}
\tablecolumns{8}
\tablewidth{0pt}
\tablecaption{Parameter space explored of both binaries modeled in
  this work.} 
  \tablehead{
    \colhead{System} &
    \colhead{r$_{ap}$} & 
    \colhead{$e$} &
    \colhead{$a$} & 
    \colhead{r$_{peri}$} & 
    \colhead{\#} &
    \colhead{Mass} &
    \colhead{Notes} 
\\
    \colhead{} &
    \colhead{(R$_{\odot}$)} & 
    \colhead{} &
    \colhead{(R$_{\odot}$)} & 
    \colhead{(R$_{\odot}$)} & 
    \colhead{orbits} &
    \colhead{transfer} &
    \colhead{ } 
}
\startdata
$0.80+0.48$ M$_{\odot}$  & 4  & 0.10 & 3.63 & 3.27 & 1.0 & N & --- \\ 
                        & 4  & 0.20 & 3.33 & 2.66 & 1.0 & Y &  1  \\
                        & 4  & 0.25 & 3.20 & 2.40 & 3.5 & Y & --- \\
                        & 4  & 0.30 & 3.07 & 2.15 & 3.0 & Y & --- \\
                        & 4  & 0.35 & 2.96 & 1.92 & 0.5 & Y &  2  \\
\\[-2.0ex] \hline \\[-1.75ex]
$1.50+1.40$ M$_{\odot}$  & 6  & 0.15 & 5.22 & 4.43 & 4.0 & Y & --- \\
                        & 6  & 0.20 & 5.00 & 4.00 & 5.0 & Y & --- \\
                        & 6  & 0.25 & 4.80 & 3.60 & 6.0 & Y & --- \\
                        & 6  & 0.30 & 4.62 & 3.23 & 3.0 & Y &  3  
\\[-1.75ex]
\enddata
\tablecomments{1: only a few particles transferred.  2: mass transfer
  rate too large for the boundary to handle at first periastron
  passage; particles penetration.  3: mass transfer rate becomes too
  large after three orbits.  See text for more details.}
\label{tab:models}
\end{deluxetable}

%% file: tab2.tex
\begin{deluxetable}{c|ccccccc}
\tabletypesize{\scriptsize}
\tablecolumns{8}
\tablewidth{0pt}
\tablecaption{Gaussian parameters for the mass transfer and accretion
  episodes of both binaries modeled in this work.}
\tablehead{
\colhead{System} &
\colhead{$e$} &
\colhead{$A_1$} &
\colhead{$A_2$} &
\colhead{$\mu_1$} &
\colhead{$\mu_2$} &
\colhead{$\sigma_1$} &
\colhead{$\sigma_2$}
\\
\colhead{} &
\colhead{} &
\colhead{(M$_{\odot}$/yr)} &
\colhead{(M$_{\odot}$/yr)} &
\colhead{(P$_{\textrm{orb}}$)} &
\colhead{(P$_{\textrm{orb}}$)} &
\colhead{(P$_{\textrm{orb}}$)} &
\colhead{(P$_{\textrm{orb}}$)}
}
\startdata
$0.80+0.48$ M$_{\odot}$  & 0.25 & 9.61$\times$10$^{-2}$ & 9.60$\times$10$^{-2}$ & 0.60 & 0.60 & 0.049 & 0.048 \\
                        &      & 2.23$\times$10$^{-2}$ & 2.23$\times$10$^{-2}$ & 1.67 & 1.67 & 0.053 & 0.049 \\
                        &      & 7.06$\times$10$^{-2}$ & 7.17$\times$10$^{-2}$ & 2.73 & 2.73 & 0.049 & 0.049 \\
                        &      & 3.42$\times$10$^{-1}$ & 3.45$\times$10$^{-1}$ & 3.65 & 3.65 & 0.032 & 0.030 \\
\\
                        & 0.30 & 5.53$\times$10$^{-1}$ & 5.50$\times$10$^{-1}$ & 0.59 & 0.58 & 0.056 & 0.054 \\
                        &      & 3.27$\times$10$^{-1}$ & 3.24$\times$10$^{-1}$ & 1.64 & 1.64 & 0.056 & 0.054 \\
                        &      & 5.48$\times$10$^{-1}$ & 5.57$\times$10$^{-1}$ & 2.69 & 2.69 & 0.057 & 0.055 \\
\\[-2.0ex] \hline \\[-1.75ex]
$1.50+1.40$ M$_{\odot}$  & 0.15 & 7.97$\times$10$^{-6}$ & 7.97$\times$10$^{-6}$ & 0.59 & 0.59 & 0.051 & 0.051 \\
                        &      & 1.00$\times$10$^{-5}$ & 1.00$\times$10$^{-5}$ & 1.58 & 1.58 & 0.048 & 0.048 \\
                        &      & 1.29$\times$10$^{-5}$ & 1.29$\times$10$^{-5}$ & 2.58 & 2.58 & 0.058 & 0.058 \\
                        &      & 1.22$\times$10$^{-5}$ & 1.22$\times$10$^{-5}$ & 3.57 & 3.57 & 0.065 & 0.065 \\
\\
                        & 0.20 & 1.16$\times$10$^{-4}$ & 1.16$\times$10$^{-4}$ & 0.57 & 0.57 & 0.049 & 0.049 \\
                        &      & 1.22$\times$10$^{-4}$ & 1.29$\times$10$^{-4}$ & 1.55 & 1.55 & 0.066 & 0.071 \\
                        &      & 1.26$\times$10$^{-4}$ & 1.33$\times$10$^{-4}$ & 2.55 & 2.56 & 0.062 & 0.063 \\
                        &      & 1.44$\times$10$^{-4}$ & 1.52$\times$10$^{-4}$ & 3.56 & 3.56 & 0.062 & 0.064 \\
                        &      & 1.38$\times$10$^{-4}$ & 1.48$\times$10$^{-4}$ & 4.56 & 4.56 & 0.065 & 0.067 \\
\\
                        & 0.25 & 1.80$\times$10$^{-3}$ & 1.80$\times$10$^{-3}$ & 0.57 & 0.57 & 0.041 & 0.041 \\
                        &      & 2.27$\times$10$^{-3}$ & 2.29$\times$10$^{-3}$ & 1.57 & 1.57 & 0.043 & 0.043 \\
                        &      & 2.79$\times$10$^{-3}$ & 2.85$\times$10$^{-3}$ & 2.57 & 2.57 & 0.045 & 0.047 \\
                        &      & 3.35$\times$10$^{-3}$ & 3.46$\times$10$^{-3}$ & 3.57 & 3.57 & 0.045 & 0.047 \\
                        &      & 3.64$\times$10$^{-3}$ & 3.79$\times$10$^{-3}$ & 4.56 & 4.57 & 0.042 & 0.045 \\
                        &      & 3.87$\times$10$^{-3}$ & 4.07$\times$10$^{-3}$ & 5.56 & 5.57 & 0.043 & 0.046 \\
\\
                        & 0.30 & 3.10$\times$10$^{-2}$ & 3.08$\times$10$^{-2}$ & 0.57 & 0.57 & 0.036 & 0.036 \\
                        &      & 3.41$\times$10$^{-2}$ & 3.47$\times$10$^{-2}$ & 1.57 & 1.57 & 0.037 & 0.037 \\
                        &      & 4.05$\times$10$^{-2}$ & 4.13$\times$10$^{-2}$ & 2.57 & 2.57 & 0.037 & 0.037 
\\[-1.75ex]
\enddata
\label{tab:gaussfits}
\end{deluxetable}

%% file: ms.bbl
\begin{thebibliography}{}
\bibitem[Bate et al.(1995)]{bate1995} Bate, M.R., Bonnell, I.A., \& Price, N.M. 1995, \mnras, 277, 362
\bibitem[Benz(1990)]{benz1990} Benz, W. 1990, in Buchler J.R., ed. The Numerical Modeling of Nonlinear Stellar Pulsations: Problems and Prospects. Kluwer, Dordrecht, p.26 9
\bibitem[Church et al.(2009)] {church2009} Church, R.P., Dischler, J., Davies, M.B., Tout, C.A., Adams, T., \& Beer, M.E. 2009, \mnras, 395, 1127
\bibitem[D'Souza et al.(2006)] {d'souza2006} D'Souza, M.C.R., Motl, P.M., Tohline, J.E., \& Frank, J. 2006, \apj, 643, 381
\bibitem[Dermine et al.(2009)] {dermine2009} Dermine, T., Jorissen, A., Siess, L., \& Frankowski, A. 2009, \aap, 507, 891
\bibitem[Deupree \& Karakas(2005)] {deupree2005} Deupree, R.G., \& Karakas, A.I. 2005, \apj, 633, 418
\bibitem[Edwards \& Pringle(1987)] {edwards1987} Edwards, D.A., \& Pringle, J.E. 1987, \mnras, 229, 383
\bibitem[Eggleton(1983)]{eggleton1983} Eggleton, P.P. 1983, \apj, 268, 368     
\bibitem[Eggleton(2006)]{eggletonbook} Eggleton, P.P. 2006, Evolutionary Processes in Binary and Multiple Stars. Cambridge Univ. Press, Cambridge
\bibitem[Gokhale et al.(2007)]{gokhale2007} Gokhale, V., Peng, X.M., \& Frank, J. 2007, \apj, 655, 1010
\bibitem[Guenther et al.(1992)]{guenther1992} Guenther, D.B., Demarque, P., Kim, Y.-C., \& Pinsonneault, M.H. 1992, \apj, 387, 372
\bibitem[Han, Tout, \& Eggleton(2000)]{han2000} Han, Z., Tout, C.A., \& Eggleton, P.P. 2000, \mnras, 319, 215
\bibitem[Hurley et al.(2002)]{hurley2002} Hurley, J.R., Tout, C.A., \& Pols, O.R. 2002, \mnras, 329, 897
\bibitem[Iben(1991)] {iben1991} Iben, I. Jr 1991, \apjs, 76, 55
\bibitem[Iben \& Livio(1993)] {iben1993} Iben, I. Jr, \& Livio, M. 1993, \pasp, 105, 1373
\bibitem[Ivanova et al.(2005)] {ivanova2005} Ivanova, N., Belczynski, K., Fregeau, J.M., \& Rasio, F.A. 2005, \mnras, 358, 572
\bibitem[Lajoie \& Sills(2010)]{lajoie2010I} Lajoie, C.-P., \& A. Sills, 2010, \apj, submitted
\bibitem[Layton et al.(1998)] {layton1998} Layton, J.T., Blondin, J.M., Owen, M.P., \& Stevens, I.R. 1998, \na, 3, 111
\bibitem[Lombardi et al.(1999)]{lombardi1999} Lombardi, J.C. Jr., Sills, A., Rasio, F.A., \& Shapiro, S.L. 1999, \jocp, 152, 687
\bibitem[Lombardi et al.(2006)]{lombardi2006} Lombardi, J.C. Jr., Proulx, X.F., Dooley, K.L., Theriault, E.M., Ivanova, N., \& Rasio, F.A. 2006, \apj, 640, 441
\bibitem[Monaghan(1989)]{monaghan1989} Monaghan, J.J. 1989, \jocp, 82, 1
\bibitem[Morton(1960)] {morton1960} Morton, D.C. 1960, \apj, 132, 146
\bibitem[Paczy\'nski(1965)]{pacz1965} Paczy\'nski, B. 1965, \actaa, 15(2), 89
\bibitem[Paczy\'nski(1971)]{pacz1971} Paczy\'nski, B. 1971, \araa, 9, 183
\bibitem[Paczy\'nski \& Sienkiewicz(1972)]{pacz1972} Paczy\'nski, B., \& Sienkiewicz, R. 1972, \actaa, 22, 73
\bibitem[Petrova \& Orlov(1999)] {petrova1999} Petrova, A.V., \& Orlov, V.V. 1999, \aj, 117, 587
\bibitem[Portegies Zwart \& Verbunt(1996)]{pzv1996} Portegies Zwart, S.F., \& Verbunt, F. 1996, \aap, 309, 179
\bibitem[Press et al.(1992)]{press1992} Press, W.H., Teukolsky, S.A., Vetterling, W.T., \& Flannery, B.P. 1992, Numerical Recipes in Fortran, Cambridge Univ. Press, Cambridge
\bibitem[Price(2007)] {splash2007} Price, D.J. 2007, Publ. Astron. Soc. Aust., 24, 159
\bibitem[Raguzova \& Popov(2005)]{raguzova2005} Raguzova, N.V. \& Popov, S.B. 2005, Astron. Astrophys. Trans., 24, 151
\bibitem[Reg\"os et al.(2005)]{regos2005} Reg\"os E., Bailey, V.C., \& Mardling, R. 2005, \mnras, 358, 544
\bibitem[Ritter(1988)]{ritter1988} Ritter, H. 1988, \aap, 202, 93
\bibitem[Rosswog et al.(2004)]{rosswog2004} Rosswog, S., Speith, R., \& Wynn, G.A. 2004, \mnras, 351, 1121
\bibitem[Sepinsky et al.(2007a)]{sepinsky2007a} Sepinsky, J.F., Willems, B., \& Kalogera, V. 2007a, \apj, 660, 1624
\bibitem[Sepinsky et al.(2007b)]{sepinsky2007b} Sepinsky, J.F., Willems, B., Kalogera, V., \& Rasio, F.A. 2007b, \apj, 667, 1170 (SWKR07)
\bibitem[Sepinsky et al.(2009)] {sepinsky2009} Sepinsky, J.F., Willems, B., Kalogera, V., \& Rasio, F.A. 2009, \apj, 702, 1387 (SWKR09)
\bibitem[Sills \& Lombardi(1997)] {sills1997} Sills, A.I., \& Lombardi, J.C.Jr 1997, \apj, 484, L51
\end{thebibliography}
